A Network Analysis Approach to fMRI Condition-Specific Functional Connectivity

Svetlana V. Shinkareva[1*], Vladimir Gudkov[2], Jing Wang[1]

[1] Department of Psychology
[2] Department of Physics
University of South Carolina

*Corresponding author:
Dr. Svetlana V. Shinkareva
Department of Psychology
University of South Carolina
Columbia, SC, 29208
Phone: (803) 777 6189
Fax: (803) 777 9558
Email: shinkareva@sc.edu

Abstract

In this work we focus on examination and comparison of whole-brain functional connectivity patterns measured with fMRI across experimental conditions. Direct examination and comparison of condition-specific matrices is challenging due to the large number of elements in a connectivity matrix. We present a framework that uses network analysis to describe condition-specific functional connectivity. Treating the brain as a complex system in terms of a network, we extract the most relevant connectivity information by partitioning each network into clusters representing functionally connected brain regions. Extracted clusters are used as features for predicting experimental condition in a new data set. The approach is illustrated on fMRI data examining functional connectivity patterns during processing of abstract and concrete concepts. Topological (brain regions) and functional (level of connectivity and information flow) systematic differences in the ROI-based functional networks were identified across participants for concrete and abstract concepts. These differences were sufficient for classification of previously unseen connectivity matrices as abstract or concrete based on training data derived from other people.

Introduction

Functional Magnetic Resonance Imaging (fMRI) data simultaneously provide information about anatomical specificity of cognitive processes (functional specialization) and interactions among brain regions (functional integration). Typically, fMRI data is analyzed using statistical parametric method, with the goal of detecting differences in activation between conditions. A complementary approach to detecting differences in activation is to examine the interaction among regions, or so-called functional connectivity, defined as the temporal correlations in neural activity among distinct brain regions (Friston 1994). Functional connectivity can be analyzed either with model-based (e.g., seed-based approaches) or data-driven (e.g., Independent Component Analyses, network analysis) approaches. The later do not require *a priori* knowledge of the functional organization in the brain. Graph theoretical analysis, or network analysis, is a data-driven approach to study large scale networks in the brain, and it has been successfully applied to neuroimaging data (See Bullmore and Sporns (2009), and Stam and Reijneveld (2007) for overviews of the approach and applications to neuroimaging). For network analysis, non-overlapping brain regions (voxels or regions of interest) are represented by nodes, and the strength of association between different brain regions is represented by edges. This representation allows describing the functional connectivity in the whole brain in terms of network statistics (e.g., density) and makes it possible to identify groups of highly interconnected regions, or *communities*, in the network. The community structure of the brain networks has been examined in resting state functional MRI data (Ferrarini, et al. 2009; Meunier, et al. 2009a; Meunier, et al. 2009b; Shen, et al. 2010), functional data in rodents (Schwarz, et al. 2009), and structural connectivity data (Chen, et al. 2008).

While most network analysis studies focused on resting state connectivity (e.g., Achard and Bullmore 2007; Dosenbach, et al. 2007), functional connectivity can also be examined for each of the experimental conditions or distinct states of a cognitive task. This allows to make a comparison in functional connectivity patterns between conditions (Dodel, et al. 2005; Rissman, et al. 2004). Direct examination and comparison of condition-specific matrices is challenging due to the large number of elements in a connectivity matrix. Even for a coarse parcellation of the brain into 84 anatomical regions there are 3,486 unique elements of the connectivity matrix to consider for an undirected graph. The goal of this paper is to examine and compare whole-brain functional connectivity patterns across experimental conditions using network analysis. We achieve this goal by (1) describing whole-brain condition-specific connectivity in terms of groups of highly interconnected brain regions that are specific to an experimental condition, thus focusing on the most salient features of the connectivity matrix, and (2) identification of connectivity patterns in previously unseen data, thus testing the consistency of extracted condition-specific networks across people.

To describe whole-brain functional connectivity patterns under different experimental conditions we introduce a novel application of a cluster extraction algorithm from physics to neuroimaging data. For each participant's data set we construct condition-specific functional networks. Treating brain as a complex system in terms of a network and applying dynamical simplex evolution (DSE) method (Gudkov, et al. 2008), we partition each condition-specific network into groups of nodes (clusters) that are more tightly linked, and, at the same time, weakly linked to nodes outside the group, thus extracting the most relevant connectivity information out of potentially large number of connections. The DSE algorithm has been shown to be very efficient in comparison to other community detection methods based on its resolution power and moderate computational complexity (Gudkov, et al. 2008). To test the consistency of extracted networks for each of the conditions across people we use extracted clusters of the obtained network, which represent functionally connected regions, as features in building classifier models for identification of condition-specific networks in previously unseen data.

We illustrate our approach on abstract and concrete concept representation. How abstract and concrete concepts are represented in the brain is relevant to understanding language function in both healthy and clinical populations (Eviatar, et al. 1990; Kuperberg, et al. 2008; Mervis and John 2008). Behavioral differences in processing between abstract and concrete concepts have been well documented and referred to as the concreteness effect: concrete words are acquired earlier, and are remembered and recognized more rapidly than abstract words (Kroll and Merves 1986; Schwanenflugel 1991). A reverse concreteness effect (Warrington 1975), showing more severe impairment for understanding concrete than abstract concepts, has been found in patients with semantic dementia (Breedin, et al. 1994; Reilly, et al. 2006). This double dissociation suggests a difference in neural representation of abstract and concrete concepts. Indeed, neuroimaging evidence suggests greater engagement of the verbal system for processing of abstract concepts and greater engagement of the perceptual system for processing of concrete concepts, likely via mental imagery (Binder, et al. 2009; Wang, et al. 2010). Most of the neuroimaging studies to date investigating representation of abstract and concrete concepts examined differences in activation. One study used Dynamic Causal Modeling to investigate functional connectivity during processing of action-related and abstract concepts using volumes of interest (Ghio and Tettamanti 2010). In this work we go beyond the activation studies and examine how whole brain (large scale brain networks) functional connectivity patterns are modulated by experimental condition: making semantic similarity judgments on either abstract or concrete words.

We first present a general approach for examination and comparison of whole-brain functional connectivity patterns under different experimental conditions, and then illustrate it on a specific example of semantic representation with two experimental conditions: making semantic similarity judgments on either abstract or concrete words.

## Methods

*Condition-specific functional connectivity*

Condition-specific functional connectivity between preprocessed time series $x = (x_1, ..., x_m)^T$ and $y = (y_1, ..., y_m)^T$ for each pair of brain regions can be estimated using weighted correlation coefficient (Dodel, et al. 2005):

$$c_\rho(x, y) = \frac{\text{cov}_\rho(x, y)}{\sqrt{\text{var}_\rho(x) \text{var}_\rho(y)}},$$

where, $\text{cov}_\rho(x, y) = \sum_{t=1}^{m} \rho_t x_t y_t$, m is the number of time points, $\text{var}_\rho(x) = \text{cov}_\rho(x,x)$, and the weight function $\rho = (\rho_1, ..., \rho_m)^T$ is constructed by convolving the vector of onsets for each of the conditions (e.g., abstract and concrete) with the canonical hemodynamic response function. To ensure the real-valued correlation all the values in the weight vector are made positive by taking the absolute value (Dodel, et al. 2005). The difference(s) between condition-specific functional connectivity matrices can be examined directly (Dodel, et al. 2005). In cases when the fixation condition is also available, a baseline connectivity matrix can be constructed based on fixation condition. To eliminate background fluctuations unrelated to condition-specific neural processes (e.g., physiological artifacts) the difference between functional connectivity during fixation and each of the conditions can be examined. To focus on strongest functional connections, these condition-specific connectivity matrices can be thresholded, and treated as undirected weighted adjacency matrices for network analysis methods (Figure 1A).

*Network analysis: cluster identification*

Condition-specific connectivity matrices contain a lot of information; the number of unique elements in undirected graph can easily reach thousands (n(n-1)/2 in a network with n nodes). Our goal is to extract the most informative subset of nodes, or main topological structure of a network. To focus on the most informative subset of the nodes and to reduce the dimensionality of the data we use DSE method to reconstruct the connectivity matrices in terms of groups of highly interconnected nodes. The main idea of the algorithm is as follows. Consider a network with *n* nodes represented by an *n × n* connectivity matrix *C*, such that $C_{ij} = 1$ for connected nodes *i* and *j*, and $C_{ij} = 0$ if there is no link between them. In general, elements of $C_{ij}$ can have different values describing the nature and intensity of connections. A single network can be represented by many matrices which are related by a similarity transformation $C = U^{-1}CU$, where $U$ $(U^{-1} = U^\dagger)$ is an *n×n* matrix affecting the permutation of rows and columns (many possible orderings of brain regions in a matrix). Let the *n* nodes be represented by *n* point masses in an *n-1* dimensional space at locations $\vec{r}_i(\tau)$, *i* = 1, ..., *n*. At τ=0 the $\vec{r}_i(\tau)$ are placed in a completely symmetric and unbiased manner at the *n* vertices of a symmetric simplex inscribed inside the unit sphere in *n*-1 dimensions so that $\vec{r}_i^2 = 1$ holds for all vertices. Thus the distance between any pair of vertices of the starting simplex of the dynamical simulation is:

$$|\vec{r}_i - \vec{r}_j| = \sqrt{\frac{2n}{n-1}}$$

The forces acting on the nodes are introduced to represent the links between nodes. These forces could be either attractive or repulsive according to whether the pair of nodes are connected or disconnected. Therefore, the vertices have the tendency to move towards each other to form clusters. The dynamics of the vertices are governed by the forces defined from the connectivity matrix and the vertex displacements vary from vertex to vertex according to their mutual connectivities. Thus, after a small number of steps the new vertex positions accurately depict the cluster structure of the network, since the mutual distances between vertices in the same cluster are systematically smaller than the distances between vertices from different clusters. Thus, clusters are identified by choosing an adequate maximum threshold for the mutual distances in the *n-1* dimensional space, and arranging the nodes in groups that correspond to neighborhoods formed with vertices with mutual distances smaller than the chosen threshold.

More formally, let $\vec{F}_{ij}$ denote "forces" between point masses in the *n*-1 dimensional space corresponding to connected pairs of nodes in the network. These forces group strongly interconnected nodes by moving them to spatial proximity. As the forces displace the point masses representing the nodes from their original symmetric positions the mutual distances $|\vec{r}_i - \vec{r}_j|$ keep changing and the simplex evolves. When $C_{ij} \neq 0$ the force between the point masses *i* and *j* acts in the direction of $\vec{r}_i - \vec{r}_j$ as:

$$\vec{F}_{ij} = C_{ij} f(|\vec{r}_i - \vec{r}_j|) \frac{\vec{r}_i - \vec{r}_j}{|\vec{r}_i - \vec{r}_j|},$$

where *f(r)* is the same for all connected pairs. Here, we use *f(r)* = 1. To avoid "overshoots" and oscillations the point masses move according to "Aristotelian Dynamics", simulating the motion of the vertices as the motion of point masses in a liquid with a high viscosity $\mu_i$:

$$\mu_i \frac{d\vec{r}_i}{d\tau} = \vec{F}_i,$$

where $\vec{F}_i = \sum_{j \neq i} \vec{F}_{ij}$ (we use $\mu_i$ =1). Using time increments $\delta$ the position of a node at each step of the algorithm is given by:

$$\vec{r}_i(\tau + \delta) = \vec{r}_i(\tau) + \frac{\delta}{\mu_i} \vec{F}_i(\vec{r}_i(\tau)), \text{ } i=1,\ldots, n.$$

The mutual distances $d_{ij} = |\vec{r}_i - \vec{r}_j|$ between each pair of nodes are calculated after each step of the algorithm. To identify the communities a maximum threshold for the mutual distances is chosen in the $n$-1 dimensional space. Then, for a given threshold $\varepsilon$, two nodes *i* and *j* belong to the same cluster if $|\vec{r}_i - \vec{r}_j| < \varepsilon$. Thus, the procedure groups the vertices according to how tightly the nodes are connected to each other. At each step of the algorithm one can apply a set of different thresholds which provides instant (spectroscopic) resolution for the cluster and all sub-cluster structure of the network. The important point of the approach is that the forces are governed by the elements of the adjacency matrix, which can represent the specific relations between nodes such as frequency of communication, intensity of the information exchange, or any other parameter to describe the level of connection. Therefore, different representations for the adjacency matrix of the given network result in different topological structures. This gives the opportunity to describe networks in terms of both structural (based on the connectivity of nodes) topology as well as a dynamical topology based on specific properties of nodes communication and information exchange.

To select the number of algorithm steps we use entropy, a characteristic of the system state that contains information related to the general organization of the system. In addition to node properties (e.g., degree of connectivity), it takes into account parameters of interactions between each pair of nodes (e.g., information exchange). Thus, the organization (or disorganization) of the network can be described in terms of mutual entropy. Entropy can be defined as $S = -\sum_{k=1}^{n} p_k \log_2 p_k$, where the "probabilities" $p_k = \sum_{i=1}^{n} C_{ik} / \sum_{i,j=1}^{n} C_{ij}$ form vector *P* for each node *k* (Gudkov and Montealegre 2008). Larger $p_k$ indicate that node k is more connected. Then, mutual information of a network (which is equal to negative mutual entropy) can be written as:

$$I(C) = I(P(row)) + I(P(column)) - I(P(column) | P(row)),$$

where for the Shannon case $I(P) = -S$ and

$$I(P(column) | P(row)) = \sum_{i,j}^{n} C_{ij} \log(C_{ij}).$$

Then, calculating the mutual entropy of the network at each step the algorithm stops when the entropy reaches its minimal value since this corresponds to the most organized structure of clusters on the given network (see Figure 1B). The output of the DSE calculations is mutual distances between nodes (Figure 1C). The smaller distances correspond to more connected nodes of node clusters. The cluster(s) of nodes (Figure 1D) are extracted from initially "non-organized" network (Figure 1A) by choosing an appropriate threshold for mutual distances. Thus whole-brain functional connectivity matrices are described in terms of groups of highly interconnected brain regions, focusing on the most salient features of the connectivity matrix.

*Classification*

To test the consistency of extracted condition-specific networks across participants we describe a classification procedure for identification of connectivity patterns in previously unseen data. A classification procedure is presented for two conditions. A classifier is trained on data from (*N*-1) participants to label the connectivity matrices for the left out participant. This process is repeated iteratively leaving out each of the participants (leave-one-participant-out cross validation).

Feature selection is conducted by combining the data of all participants except the one left out in a cross-validation step. For each condition, a weighted average matrix across participants in the training data is computed, weighting the participant's connectivity matrices $G_i$ by how similar they are to each other (Abdi, et al. 2009; Shinkareva, et al. 2006). Each weighted average matrix is constructed such that participants with similar connectivity patterns to those of other participants are assigned larger weights, and participants with connectivity patterns most different from those of others are assigned lower weights. The weights are given by the elements of the first eigenvector of between-participant similarity matrix *B*, rescaled to sum up to one. The *ik*-elements of *B* are computed by an RV-coefficient (Escoufier 1973; Robert and Escoufier 1976):

$$B_{ik} = \frac{tr(G_i^T G_k)}{\sqrt{tr(G_i^T G_i) \cdot tr(G_k^T G_k)}}$$

Features (elements of the connectivity matrices) used for classification are selected automatically using the DSE method from the weighted condition-specific connectivity matrices computed based on the training data. Thus out of all elements in the connectivity matrix a union of clusters extracted with the DSE method for the condition-specific networks is selected. Since the matrices are symmetric, only half of the connectivity matrix is used to avoid counting connections twice.

Classification of connectivity matrices for the left out participant is done by selecting a pairing with the best similarity score. The similarity score *β(A)* between the training $A_{train}$ and the test $A_{test}$ connectivity matrices can be computed as a cosine similarity between the two vectors constructed from these matrixes by adjusting rows one-by-one into vectors which belong to n-dimensional linear space, which can be calculated for symmetric matrices as :

$$\beta(A) = \frac{tr(A_{train}^T \cdot A_{test})}{\sqrt{tr(A_{train}^T \cdot A_{train}) \cdot tr(A_{test}^T \cdot A_{test})}}.$$

When vectors are parallel (*β(A)=1*) the two matrices are identical, and when vectors are orthogonal (*β(A)=0*) the two matrices are maximally different. Similarity match score for the candidate pairing is computed as a sum of the two cosine similarities (Mitchell, et al. 2008) for conditions *A* and *C*:

$$\beta(A \cup C) = \beta(A) + \beta(C)$$

Classification performance can be evaluated by applying the classifier to the data from left out participant. Classification accuracies are then averaged across the participants.

The expected chance accuracy of a two class classification for equal number of samples in each class is 0.5. To determine p-values for the observed classification accuracies to reject the null hypothesis that classification performance is at chance, observed classification accuracies can be compared to the empirically derived chance distribution (constructed separately for each threshold). At each fold, 1000 random permutations of the nodes are generated, and elements in training matrices are reordered based on the permutations. Cosine similarities are computed for test data and reordered training data based on the features derived from unordered training data (cluster extraction does not depend on row ordering).

*fMRI data example: abstract and concrete concepts*

Thirteen right-handed healthy volunteers from the University of South Carolina community gave written informed consent approved by the University of South Carolina Institutional Review Board and participated in the study.

Stimuli included quadruples of semantically synonymous written words from two abstract (*cognition*, *emotion*) and a two concrete (*tools*, *dwellings*) categories, with four exemplars per category (Table A1). Each quadruple (e.g., *error*, *mistake*, *blunder*, *slipup*) represented the same underlying concept (e.g., *error*). For each presentation of an exemplar, three words out of a quadruple were shown in two rows: one above and two below (Figure A1). Repetitions of the same exemplar consisted of different subsets of triplets from the four words (taking row positioning into consideration 12 unique triplets exist for each quadruple of words), with each triplet shown only once. Triplets from abstract and concrete categories were equated on word length and frequency.

While being scanned, the participants performed a semantic similarity judgment task (Breedin, et al. 1994; Noppeney and Price 2004; Sabsevitz, et al. 2005) during which they decided which word of the two words at the bottom of the display is more similar to a third word at the top of the display. Participants indicated their choice with a button press of an index or a middle finger on their right hand. This task was selected to prompt careful evaluation of each item and its properties, thus covertly eliciting concept representation for a prolonged time period (3 s).

Each word triple was presented for 3 s, followed by 7 s rest period, during which participants were instructed to fixate on an X displayed in the center of the screen (Figure A1). There were 6 additional presentations of fixation, 24 s each, distributed across the session to provide a baseline measure of brain activation. Each exemplar was presented 6 times during the experiment, thus a total of 48 concrete and 48 abstract triplets were shown. This experimental procedure has been designed for multi-voxel pattern analysis methods and was used successfully in our previous investigations of concept representation (Shinkareva, et al. 2008).

Functional images were acquired on a Siemens Trio 3.0T scanner with a twelve-channel headcoil at the McCausland Center for Brain Imaging at the University of South Carolina using a gradient echo EPI with TR=2.2 ms, TE=30 ms and 90° flip angle. Thirty-six 3 mm thick axial slices were imaged in ascending order with a gap of 0.3 mm between slices. The acquisition matrix was 64x64 with 3x3x3 $mm^3$ voxels.

The preprocessing was carried out using Statistical Parametric Mapping software (SPM5, Wellcome Department of Cognitive Neurology, London, UK). The data were corrected for motion, linear trend, and temporally smoothed with a high-pass filter using a 128 s cutoff. So that brain regions can be compared systematically across participants, the data were normalized to the MNI template brain image using a 12-parameter affine transformation and resampled to 3x3x3 $mm^3$ voxels.

For simplicity, the total number of nodes was reduced. Forty-two pairs of cortical and sub-cortical regions of interest (ROI) were defined anatomically (Achard and Bullmore 2007; Liu, et al. 2008) using Anatomical Automatic Labeling (AAL) system (Tzourio-Mazoyer, et al. 2002). An average time series was extracted for each of the 84 regions.

Two condition-specific connectivity matrices were constructed for each participant: one for abstract concepts and one for concrete concepts. These 26 matrices (13 participants x 2 conditions) were used for density computation and cross-participant classification.

Results

Because there is no one preferred way to choose the threshold, we have studied abstract and concrete condition-specific matrices at 12 threshold levels: 0.75, .80, 0.85, 0.90, 0.95 and 0.99 percentile of the correlations distribution for positive correlations and 0.01, 0.05, 0.10, 0.15, 0.20, and 0.25 percentile of the correlations distribution for negative correlations. Network analysis revealed different functional connectivity patterns during processing of abstract and concrete words. These differences in the functional networks of brain activity were both topological (brain regions) and functional (functional connectivity) and were consistent across participants.

For positive correlations, network analyses identified one cluster of positively correlated brain regions that were highly connected during the processing of abstract words, and, separately, one cluster of brain regions that were highly connected during the processing of concrete words. The clusters of tightly linked brain regions identified by the network analysis for abstract and concrete networks consisted of largely overlapping sets of brain regions located in frontal, temporal, parietal as well as limbic lobes. Although the nodes in the two extracted networks were about the same, the connectivity patterns among them were different. For positive correlations, there was a tendency for higher density (proportion of connections present out of all possible connections) of the networks for concrete words than for abstract words. This finding was consistent across different threshold levels (Figure 2). For illustration, the network clusters for abstract and concrete concepts extracted from the weighted average connectivity matrices computed across all participants with 0.95 threshold (top 5% of positive correlations) are shown in Figure 3. From 84 brain regions, the DSE algorithm identified 30 regions for abstract and 31 regions for concrete networks that are relevant to language tasks and have been previously implicated in semantic processing (Bookheimer 2002).

For negative correlations, network analyses identified one cluster of negatively correlated brain regions that were highly connected during the processing of abstract words, and, separately, one cluster of brain regions that were highly connected during the processing of concrete words. Similarly to positive correlations, there was a tendency for higher density of the networks for concrete words than for abstract words. This finding was consistent across different threshold levels (Figure 4). The two networks derived from the weighted average connectivity matrices computed across all participants with 0.05 threshold (top 5% of negative correlations, in magnitude) are shown in Figure 5.

Having described the differences in the whole brain functional connectivity for abstract and concrete conditions across participants we have checked if those differences were sufficient to accurately identify unlabeled networks from a previously unseen participant. Data from all but one participant were used to identify the two connectivity networks derived from the left-out participant. To determine if it was possible to identify one of two connectivity matrices as abstract or concrete in the left-out participant using brain activation data from other participants, a classifier was trained on the network data combined from all but one participant. This procedure was repeated for each of the participants, and mean classification accuracies were reported across all participants. We were able to label the two networks with accuracies above chance level (Tables 1 and 3). For example, we identified networks with 0.77 accuracy (p-value = 0.01) when training on the top 5% of positive correlations and with 0.85 accuracy (p-value = 0.00) when training on the top 25%, in magnitude, of negative correlations. For comparison, classification accuracy was 0.54 with no feature selection, using n(n-1)/2 elements of the connectivity matrix. Classification accuracies based on elements of the connectivity matrices (with no additional feature selection) surpassing the thresholds of 0.75, .80, 0.85, 0.90, 0.95 and 0.99 percentile of the correlations distribution for positive correlations were not significant, and are shown in Tables 2. Classification accuracies based on elements of 0.01, 0.05, 0.10, 0.15, 0.20, and 0.25 percentile of the correlations distribution for negative correlations were comparable to the accuracies of using DSE algorithm, and are shown in Tables 4. These findings indicate that (1) there are systematic differences in functional connectivity between abstract and concrete conditions; and (2) a high degree of commonality across participants exist in the connectivity patterns elicited by abstract and concrete words. This commonality in connectivity patterns enabled the cross-participant condition-specific network identification as abstract or concrete.

## Discussion and Conclusions

We have presented a general framework to describe whole brain functional connectivity patterns across experimental conditions using network analysis. It allows to simultaneously recover topological (brain regions) and functional (functional connectivity) structure of the given network using the strength of association among regions. The approach has several advantages. First, it is data-driven and does not require *a priori* selection of regions of interest, in contrast to seed-based connectivity methods. Second, from a large number of elements in a connectivity matrix it extracts a smaller number of highly interconnected elements. Examining clusters of highly interconnected elements is more informative compared to examining highest correlation elements in the connectivity matrix, because it provides detailed information about cluster structure including sub-clusters of a given network. This is especially important for networks with a large number of nodes, where even the identification of self- related elements in the connectivity matrix (clusters) is a very difficult and ambigious problem. The approach is used to describe differences between functional connectivity patterns of two conditions and is illustrated on an fMRI data set of different connectivity patterns during processing of abstract and concrete concepts. The algorithm is general and can be used to examine either structural or functional connectivity.

There are many different ways to convert fMRI data to an adjacency matrix with no clear preference among them (Hayasaka and Laurienti in press; Stam and Reijneveld 2007; Zalesky, et al. in press) . The choice of nodes for anatomical networks has been shown to have no effect on binary decisions, such as scale-free properties (Zalesky, et al. in press). At the same time, finest scale (voxel) resolution has been shown to have benefits for examining functional resting state fMRI networks (Hayasaka and Laurienti in press). We have defined the nodes based on AAL mask, as typical for most network analysis studies (Achard and Bullmore 2007; Achard, et al. 2006; Liu, et al. 2008; Supekar, et al. 2008), but note that the general methodology is not restricted to anatomically defined regions of interest. This method, being applied to a more detailed structure (better spatial resolution), has a potential to recover more hidden functional connectivities and, as a consequence, to provide detailed functional characteristic of brain behavior, since it was previously tested on a number of large (more than thousands nodes) different networks and showed the best resolution and performance (Gudkov, et al. 2008). Investigation of other ways to compute functional connectivity and how that would impact the results is of interest but beyond the scope of this paper.

To test the consistency of extracted condition-specific networks across people we have successfully identified networks as abstract or concrete in previously unseen data. Classification of functional connectivity patterns is related to other approaches to predicting neuroimaging results based on network structure, such as predicting resting-state functional connectivity from structural connectivity (Honey, et al. 2009) and classification of schizophrenia patients (Cecchi, et al. 2009).

Systematic differences in functional connectivity were identified between abstract and concrete conditions, complementing existing functional localization studies. The consistency of these differences across participants was highlighted by the ability to accurately label the two connectivity matrices from a previously unseen data as abstract or concrete based on data from other participants. This ability to identify one of two networks across participants reveals common connectivity patterns during processing of abstract and concrete words across people. These results are consistent with the connectivity model for semantic processing (Klimesch 1994) and the theoretical framework of representational differences between object concepts and (more abstract) relational concepts (Gentner 1981). Indeed, concrete concepts have richer context than abstract concepts (Schwanenflugel, et al. 1988). Examining network connectivity in the brain can further advance the understanding of abstract and concrete concept representation. It would be interesting to compare these results to condition-specific networks derived from data collected in another neuroimaging modality, for example MEG.

Acknowledgments

We thank Laura Bradshaw Baucom for helpful comments on the manuscript.

Figure 1: Dynamical simplex evolution algorithm illustration. (A) Thresholded connectivity matrix. (B) Optimization function for stopping DSE algorithm – algorithm stops when the maximal mutual information (minimum entropy) value is reached. (C) The output of the DSE algorithm is mutual distances between nodes. (D) Reordered connectivity matrix showing cluster of nodes identified by the DSE algorithm (projection of nodes from (C) with mutual distances below selected threshold).

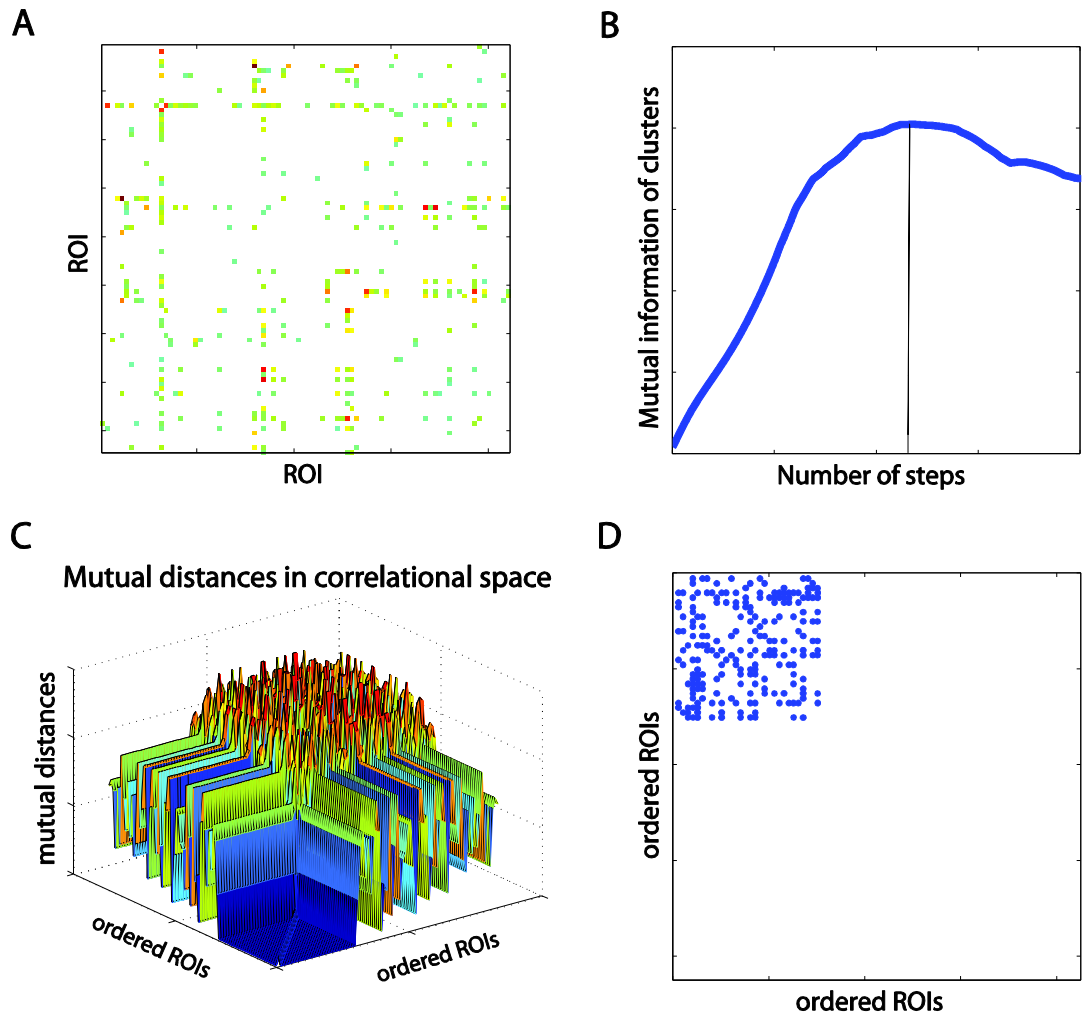

Figure 2: For positive correlations, there was a tendency for higher density of the networks for concrete words than for abstract words. The mean density across the participants and standard errors are shown for each of the two networks at multiple threshold levels.

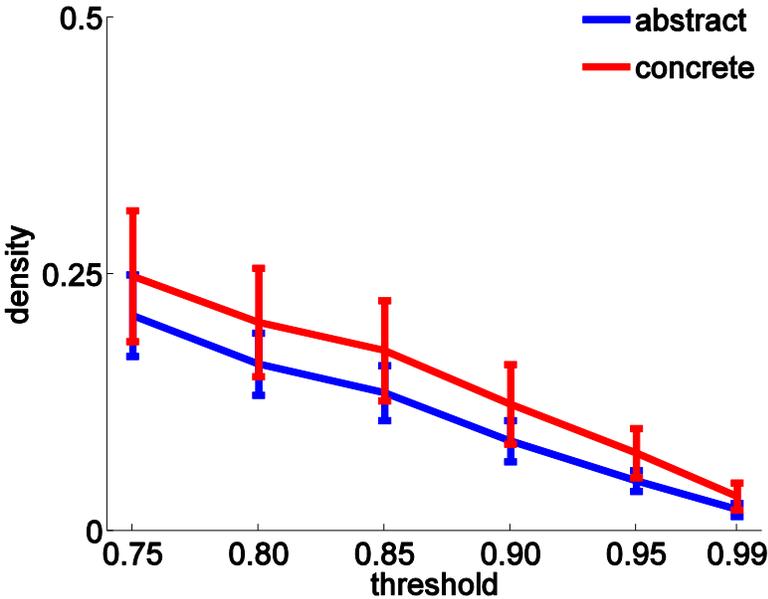

Figure 3: Extracted networks (shown unweighted) from the connectivity matrices weighted across participants and thresholded at the top 5% of positive correlations. Nodes are placed at a center of mass of each ROI.

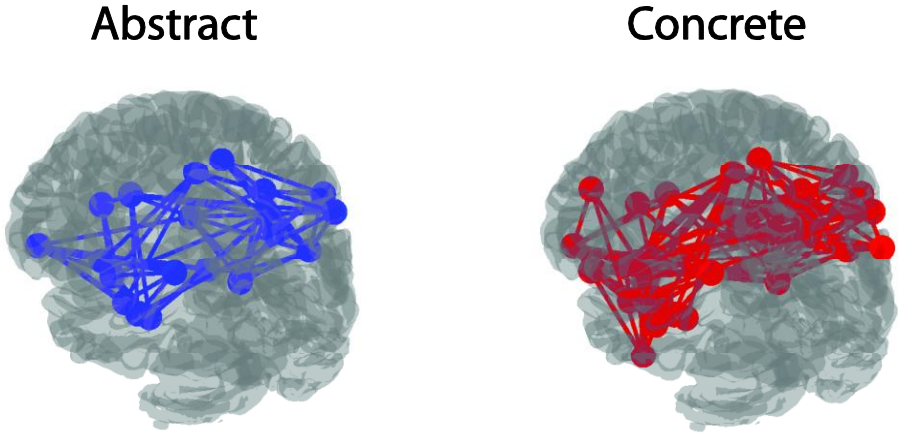

Figure 4: For negative correlations, there was a tendency for higher density of the networks for concrete words than for abstract words. The mean density across the participants and standard errors are shown for each of the two networks at multiple threshold levels.

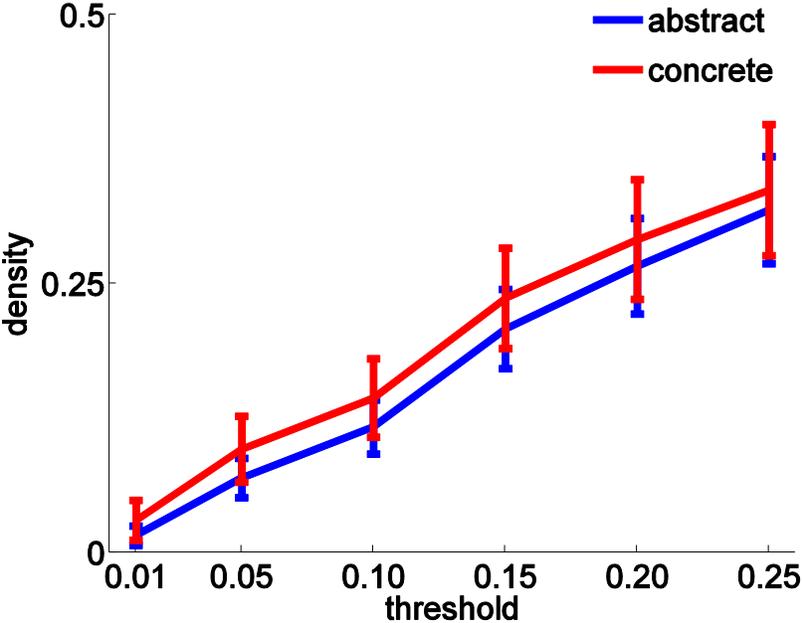

Figure 5: Extracted networks (shown unweighted) from the connectivity matrices weighted across participants based on the top 5% of negative correlations. Nodes are placed at a center of mass of each ROI.

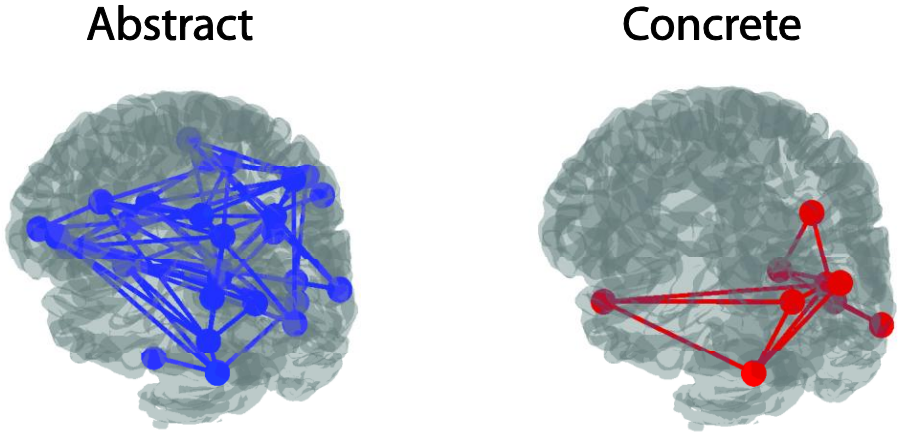

Appendix

Table A1: Stimuli quadruples by category.

Abstract (cognition)

| "error"   | error      | mistake     | blunder     | slipup     |
| "concept" | concept    | reflection  | notion      | cogitation |
| "opinion" | opinion    | contention  | conviction  | belief     |
| "illusion"| dream      | delusion    | fantasy     | illusion   |

Abstract (emotion)

| "pity"    | mercy      | empathy     | sympathy    | pity       |
| "delight" | happiness  | enjoyment   | pleasure    | delight    |
| "sadness" | sadness    | distress    | depression  | gloom      |
| "fright"  | dread      | fright      | trepidation | horror     |

Concrete (tools)

| "knife"   | knife      | blade       | scalpel     | cutlass    |
| "hammer"  | hammer     | club        | hatchet     | mallet     |
| "pencil"  | paintbrush | highlighter | pencil      | marker     |
| "fork"    | spatula    | fork        | tablespoon  | chopsticks |

Concrete (dwellings)

| "house"   | house      | apartment   | flat        | condominium|
| "palace"  | mansion    | castle      | penthouse   | palace     |
| "theatre" | stadium    | pavilion    | coliseum    | theatre    |
| "barn"    | silo       | barn        | shed        | warehouse  |

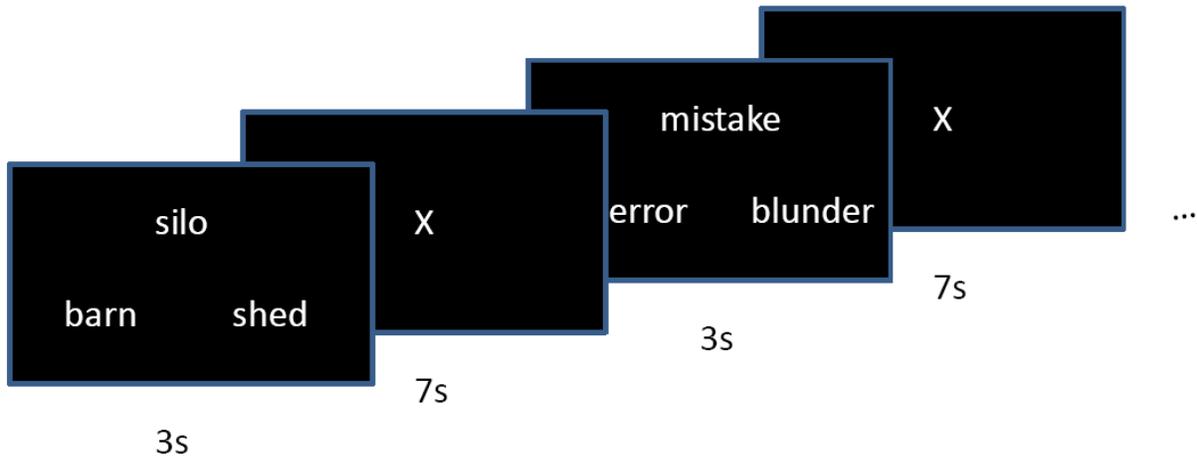

Figure A1: Schematic representation of the experimental paradigm. All stimuli were presented in white against a black background.